\documentclass[twoside,english,iop]{emulateapj}
\usepackage[T1]{fontenc}
\usepackage{graphicx}

\makeatletter

\providecommand{\tabularnewline}{\\}

\usepackage{apjfonts}
\usepackage[hyperfootnotes=false,colorlinks=true,citecolor=blue]{hyperref}

\shorttitle{TURBULENT ISM}
\shortauthors{SEON}

\makeatother

\usepackage{babel}
\begin{document}

\title{THE COLUMN DENSITY VARIANCE IN TURBULENT INTERSTELLAR MEDIA: A FRACTAL
MODEL APPROACH}

\author{Kwang-Il Seon\altaffilmark{1,2}}

\altaffiltext{1}{Korea Astronomy and Space Science Institute, Daejeon 305-348, Republic of Korea; kiseon@kasi.re.kr}
\altaffiltext{2}{University of Science and Technology, Daejeon 305-350, Republic of Korea} 
\begin{abstract}
Fractional Brownian motion (fBm) structures are used to investigate
the dependency of column density variance ($\sigma_{\ln N}^{2}$)
in the turbulent interstellar medium on the variance of three-dimensional
density ($\sigma_{\ln\rho}^{2}$) and the power-law slope of the density
power spectrum. We provide quantitative expressions to infer the three-dimensional
density variance, which is not directly observable, from the observable
column density variance and spectral slope. We also investigate the
relationship between the column density variance and sonic Mach number
($M_{{\rm s}}$) in the hydrodynamic (HD) regime by assuming the spectral
slope and density variance as functions of sonic Mach number, as obtained
from the HD turbulence simulations. They are related by the expression
$\sigma_{\ln N}^{2}=A\sigma_{\ln\rho}^{2}=A\ln(1+b^{2}M_{{\rm s}}^{2})$,
suggested by Burkhart \& Lazarian for the magneto-hydrodynamic (MHD)
case. The proportional constant $A$ varies from $\approx0.2$ to
$\approx0.4$ in the HD regime as the turbulence forcing parameter
$b$ increases from 1/3 (purely solenoidal forcing) to 1 (purely compressive
forcing). It is also discussed that the parameter $A$ is lowered
in the presence of a magnetic field.
\end{abstract}


\section{Introduction}

A variety of observations and simulations have shown that the density
structures of the interstellar medium (ISM) are scale-free, hierarchical,
and fractal \citep[e.g., ][]{Stutzki1998,Elmegreen2004,Burkhartetal2012}.
The supersonic, compressible turbulence is likely responsible for
the complex and hierarchical density structures observed in the ISM.
The density fluctuation due to the supersonic turbulence plays a crucial
role in models of star formation rates and initial mass functions
\citep{MacLow2004,Krumholz,Hennebelle2008,Federrath2012}. Hence,
the statistical properties of density structures in the turbulent
ISM have been extensively studied in hydrodynamic (HD) and magneto-hydrodynamic
(MHD) simulations.

It is now well known that the probability distribution functions (PDFs)
of the three-dimensional (3D) densities and column densities of the
turbulent ISM are close to lognormal \citep{Vazquez94,Nordlund1999,Klessen2000,Ostriker01,Wada2001,Burkhart2012}.
The standard deviation of density ($\sigma_{\rho/\rho_{0}}$) increases
with sonic Mach number, $M_{{\rm s}}$ \citep{Nordlund1999,Ostriker01}.
This trend is expressed by $\sigma_{\rho/\rho_{0}}=bM_{{\rm s}}$,
where $\rho/\rho_{0}$ is the density ($\rho$) normalized by the
mean density ($\rho_{0}$) and $b$ is a constant of proportionality,
known as the turbulence forcing parameter. Values of $b$ ranging
from $\approx$ 0.3 to $\approx$ 1.0 have been suggested in numerical
simulations \citep{Padoan97,Passot1998,Kritsuk2007,Beetz2008}. \citet{Federrath2008,Federrath2009,Federrath2010}
have found that for the same $M_{{\rm s}}$, compressive (dilatational)
forcing leads to much larger density variance compared to solenoidal
(rotational) forcing. They showed that $b$ depends on the type of
the turbulence forcing varying from $\approx1/3$ for solenoidal forcing
to $\approx1$ for compressive forcing.

The density power spectrum is also a useful tool to characterize the
turbulent ISM. The power spectra of turbulent clouds are power laws
in form \citep{Crovisier83,Desphpande2000,Stanimirovic2001,Padoan04}.
\citet{Kim05} have investigated the dependency of the power-law slope
of the density power spectrum on $M_{{\rm s}}$ and found that the
density spectrum becomes gradually shallower as $M_{{\rm s}}$ increases
in HD turbulent media. \citet{Kowal2007} and \citet{Burkhart2010}
confirmed the flattening of the density spectra with $M_{{\rm s}}$
in MHD cases. \citet{Federrath2009} showed that compressive forcing
leads to significantly steeper density spectra than solenoidal forcing.

We note that the spectral slope of the 3D density power spectrum is
the same as that of the projected column density power spectrum \citep[e.g., ][]{Stutzki1998,Padoan04}
and thus can be directly extracted from observations. On the other
hand, the 3D density is not an observationally accessible quantity,
but instead the column density, i.e., the integral of the density
along a line of sight, is observable. There is a great deal of observational
data, which traces the ISM column density, from surveys. Some examples
are the GALFA \citep{Peek2011} and GASKAP \citep{Dickey2012} \ion{H}{1}
surveys and Columbia-Cfa \citep{Dame2001} and ThrUMMS%
\footnote{http://www.astro.ufl.edu/\textasciitilde{}peterb/research/thrumms/%
} molecular gas surveys. The best way to infer the variance of the
3D density field from observations is obviously to compare the observational
data with a large number of turbulence simulations. However, instead
of performing expensive simulations, a simpler approach to derive
the 3D density variance would be of great interest. \citet{Padoan97}
generated random density distributions with various density variances
and spectral slopes and compared the resulting synthesized extinction
maps with the observed extinction data. \citet{Bruntetal2010} developed
an interesting technique for calculating the variance of a 3D density
field using only information contained in the two dimensional projection.
The technique has been applied to the Taurus molecular cloud \citep{Brunt2010}.
More recently, \citet{Burkhart2012} found a column density variance-Mach
number relationship, using MHD simulations. The relationship closely
follows the form of the 3D density variance-Mach number relationship
but includes a scaling parameter $A$ such that 
\begin{equation}
\sigma_{\ln N}^{2}=A\ln(1+b^{2}M_{{\rm s}}^{2}),\label{eq:1}
\end{equation}
 where $\sigma_{\ln N}^{2}$ is the variance of the logarithm of the
column density $N$%
\footnote{Note that $\sigma_{\ln N/N_{0}}=\sigma_{\ln N}$, but $\sigma_{N/N_{0}}=\sigma_{N}/N_{0}$
for the mean column density $N_{0}$. The same relationships are applied
to the 3D density $\rho$.%
}.

In the present study, quantitative expressions relating the 3D density
variance, spectral slope, and column density variance are obtained
by using the fractional Brownian motion (fBm) algorithm \citep{Saupe1988,Stutzki1998,Elmegreen2002}.
A lognormal density field is obtained by generating a Gaussian random
field with a power-law spectral slope ($\gamma_{{\rm g}}$) and exponentiating
the field \citep{Elmegreen2002}. This letter is organized as follows.
In Section 2, we describe how to infer the variance of the 3D density
field from the given spectral slope and column density variance. Section
3 investigates the $\sigma_{\ln N}^{2}-M_{{\rm s}}$ relationship.
Concluding remarks are given in Section 4.

\begin{figure}[t]
\begin{centering}
\includegraphics[scale=0.4]{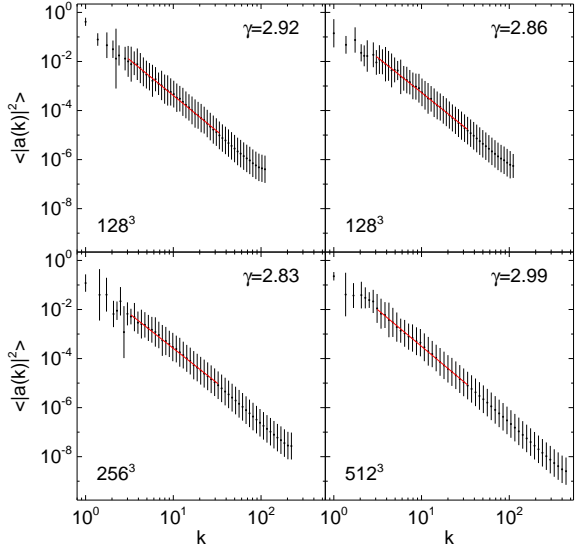}
\par\end{centering}

\caption{\label{fig1}Examples density power spectra with the 128$^{3}$, 256$^{3}$,
and 512$^{3}$ resolutions, which were obtained for ($\gamma_{{\rm g}},$
$\sigma_{\ln\rho}$) = (3.4, 1.4). Error bars denote standard deviations
in wavenumber bins. The best-fit lines and slopes, which were obtained
over the range 3 \ensuremath{\le} $k$ \ensuremath{\le} 35, are also
shown.}
\medskip{}
\end{figure}

\begin{figure}[t]
\begin{centering}
\includegraphics[scale=0.4]{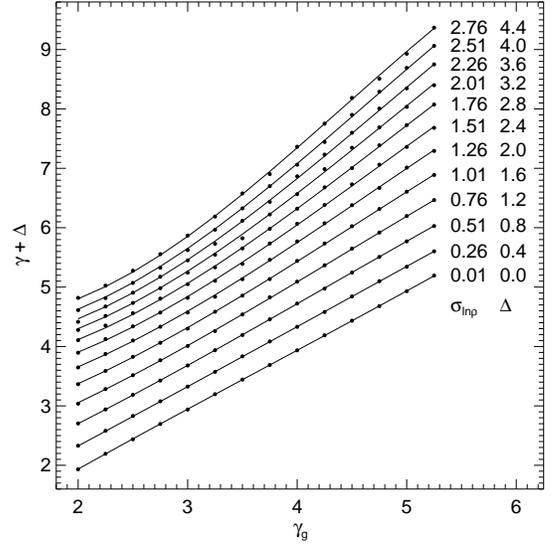}
\par\end{centering}

\caption{\label{fig2}Spectral slopes ($\gamma$) of lognormal density fields
vs. the original spectral slopes ($\gamma_{{\rm g}}$) of the Gaussian
density fields. Best-fit polynomial curves for every density variance
$\sigma_{\ln\rho}$ are also shown. The circles and curves were shifted
vertically by the amounts of $\Delta$ for clarity.}
\medskip{}
\end{figure}

\begin{figure}[t]
\begin{centering}
\includegraphics[scale=0.4]{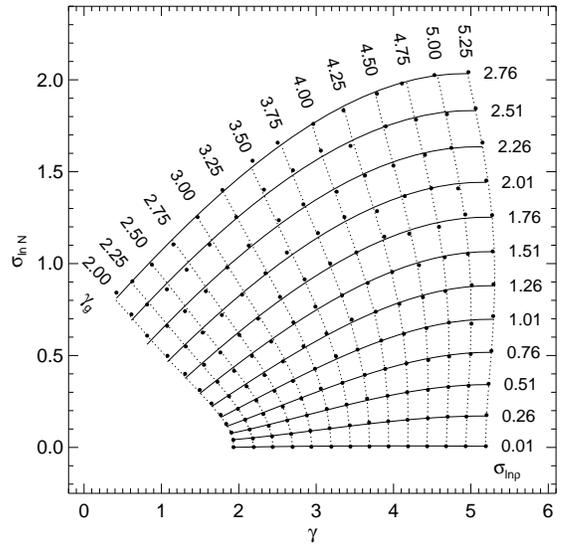}
\par\end{centering}

\caption{\label{fig3}The column density deviations $\sigma_{\ln N}$ and spectral
slopes $\gamma$ of lognormal density fields obtained for various
combinations of $\sigma_{\ln\rho}$ and $\gamma_{{\rm g}}$ of the
Gaussian random field. Solid and dotted lines are best-fit curves
tracing the ($\sigma_{\ln N}$, $\gamma$) pairs corresponding to
constant $\sigma_{\ln\rho}$ and $\gamma_{{\rm g}}$, respectively.}
\medskip{}
\end{figure}

\begin{table}[t]
\caption{\label{table}Polynomial coefficients for $\gamma$ as a function
of ($\gamma_{{\rm g}}$, $\sigma_{\ln\rho}$), and for $\gamma_{{\rm g}}$
and $\sigma_{\ln\Sigma}$ as functions of ($\gamma$, $\sigma_{\ln\rho}$).}

\begin{centering}
\begin{tabular}{cr@{\extracolsep{0pt}.}lr@{\extracolsep{0pt}.}lr@{\extracolsep{0pt}.}lr@{\extracolsep{0pt}.}lr@{\extracolsep{0pt}.}l}
\hline 
\hline & \multicolumn{2}{c}{$p_{0}$} & \multicolumn{2}{c}{$p_{1}$} & \multicolumn{2}{c}{$p_{2}$} & \multicolumn{2}{c}{$p_{3}$} & \multicolumn{2}{c}{$p_{4}$}\tabularnewline
\hline 
$a_{0}$ & $-$2&582$(-1)$ & 6&609$(-1)$ & 1&242 & $-$1&040 & 2&019$(-1)$\tabularnewline
$a_{1}$ & 1&185 & $-$6&876$(-1)$ & $-$1&019 & 7&400$(-1)$ & $-$1&462$(-1)$\tabularnewline
$a_{2}$ & $-$5&344$(-2)$ & 2&080$(-1)$ & 2&316$(-1)$ & $-$1&827$(-1)$ & 3&865$(-2)$\tabularnewline
$a_{3}$ & 4&832$(-3)$ & $-$1&928$(-2)$ & $-$1&395$(-2)$ & 1&395$(-2)$ & $-$3&283$(-3)$\tabularnewline
$b_{0}$ & 2&841$(-1)$ & $-$9&168$(-1)$ & $-$9&334$(-1)$ & 1&221 & $-$2&546$(-1)$\tabularnewline
$b_{1}$ & 8&173$(-1)$ & 5&994$(-1)$ & 1&326 & $-$1&125 & 2&119$(-1)$\tabularnewline
$b_{2}$ & 5&019$(-2)$ & $-$1&468$(-1)$ & $-$3&838$(-1)$ & 2&970$(-1)$ & $-$5&417$(-2)$\tabularnewline
$b_{3}$ & $-$4&428$(-3)$ & 1&186$(-2)$ & 3&111$(-2)$ & $-$2&375$(-2)$ & 4&329$(-3)$\tabularnewline
$c_{0}$ & $-$2&336$(-2)$ & 6&929$(-1)$ & $-$1&516 & 8&404$(-1)$ & $-$1&278$(-1)$\tabularnewline
$c_{1}$ & 1&826$(-2)$ & $-$6&199$(-1)$ & 1&099 & $-$4&586$(-1)$ & 5&987$(-2)$\tabularnewline
$c_{2}$ & $-$4&381$(-3)$ & 2&381$(-1)$ & $-$2&647$(-1)$ & 9&389$(-2)$ & $-$1&117$(-2)$\tabularnewline
$c_{3}$ & 3&689$(-4)$ & $-$2&369$(-2)$ & 2&235$(-2)$ & $-$7&710$(-3)$ & 9&542$(-4)$\tabularnewline
\hline 
\end{tabular}
\par\end{centering}

Note: The numbers in parentheses are exponents on 10, meaning that
6.609($-$1) = 6.609$\times10^{-1}$.
\end{table}

\section{MODEL OF THE LOGNORMAL DENSITY FIELD}

The fBm structures are generated by first assigning 3D Fourier coefficients
following the prescription of \citet{Elmegreen2002}. The Fourier
amplitudes are generated to be distributed as a normal Gaussian with
the variance $<\left|a(\mathbf{k})\right|^{2}>=\mathbf{|k|}^{-\gamma_{{\rm g}}}$.
The inverse Fourier transform gives a Gaussian random field $\rho_{{\rm g}}$($\mathbf{x}$).
We then multiply the density field with the desired standard deviation
of the logarithmic density $\sigma_{\ln\rho}$, and exponentiate the
field to obtain $\rho=\exp(\sigma_{\ln\rho}\rho_{{\rm g}}$).

The standard deviation $\sigma_{\ln N}$ depends on both $\gamma_{{\rm g}}$
and $\sigma_{\ln\rho}$. We should also note that the spectral slopes
$\gamma$ of resulting lognormal density fields are different from
the input slopes $\gamma_{{\rm g}}$. Therefore, we need to find the
relationship between the input variables ($\gamma_{{\rm g}}$, $\sigma_{\ln\rho}$)
and the outputs ($\gamma$, $\sigma_{\ln N}$). In addition, the generated
density fields with the same parameters ($\gamma_{{\rm g}}$, $\sigma_{\ln\rho}$)
show large fluctuations due to random phases and amplitudes in Fourier
space. We thus generated a large volume of lognormal density fields
by varying $\gamma_{{\rm g}}$ and $\sigma_{\ln\rho}$ and obtained
the average relationships between the parameters. We varied $\gamma_{{\rm g}}$
from 2.0 to 5.25 in steps of 0.25 and $\sigma_{\ln\rho}$ from 0.01
to 2.76 in steps of 0.25. One hundred random realizations with a box
size of 128$^{3}$ were generated for each combination of $\gamma_{{\rm g}}$
and $\sigma_{\ln\rho}$, resulting in a total of 16,800 realizations.
We also generated 30 realizations with a resolution of $256^{3}$
for every ($\gamma_{{\rm g}}$, $\sigma_{\ln\rho}$) pair and confirmed
that the 256$^{3}$ dataset is consistent with the results obtained
with the 128$^{3}$ resolution. As will be shown later, the present
results are also consistent with the limited samples obtained with
a box size of 512$^{3}$. The standard deviations and spectral slopes
were calculated for every realization and averaged to obtain the ensemble
average values of ($\gamma$, $\sigma_{\ln N}$) for each combination
of the input ($\gamma_{{\rm g}}$, $\sigma_{\ln\rho}$) values. The
power spectra of resulting lognormal density fields show a slight
flattening at large wavenumbers. Examples of the spectra are shown
in Figure \ref{fig1}. Spectral slopes were obtained by least-squares
fits over a wavenumber range of $3\le k\le35$. Here, the dimensionless
wavenumber is defined by $k=L/\lambda$ with the wavelength $\lambda$
and box size $L$.

For ease of use, we parameterized the relationships between the parameters
$\gamma$, $\gamma_{{\rm g}}$, $\sigma_{\ln\rho}$, and $\sigma_{\ln N}$
using polynomials. For every $\sigma_{\ln\rho}$, $\sigma_{\ln N}$
was parameterized with a cubic function of $\gamma$, i.e., $ $
\begin{equation}
\sigma_{\ln N}=\sum_{i=0}^{3}c_{i}\gamma^{i}=c_{0}+c_{1}\gamma+c_{2}\gamma^{2}+c_{3}\gamma^{3}.\label{eq:2}
\end{equation}
The coefficients $c_{i}$ ($i=0,\ldots,3$) were then fitted with
a quartic polynomial of $\sigma_{\ln\rho}$, i.e., 
\begin{equation}
c_{0}=\sum_{j=0}^{4}p_{j}\sigma_{\ln\rho}^{j}=p_{0}+p_{1}\sigma_{\ln\rho}+p_{2}\sigma_{\ln\rho}^{2}+p_{3}\sigma_{\ln\rho}^{3}+p_{4}\sigma_{\ln\rho}^{4}.\label{eq:3}
\end{equation}
In this way, $\sigma_{\ln N}$ could be estimated as a function of
$\gamma_{{\rm g}}$ and $\sigma_{\ln\rho}$. We also parameterized
$\gamma$ with a cubic function of $\gamma_{{\rm g}}$ and $\gamma_{{\rm g}}$
with another cubic function of $\gamma$ for every $\sigma_{\ln\rho}$,
i.e., $\gamma=\Sigma_{i}a_{i}\gamma_{{\rm g}}^{i}$ and $\gamma_{{\rm g}}=\Sigma_{i}b_{i}\gamma^{i}$
($i=0,\ldots,3$). The coefficients $a_{i}$ and $b_{i}$ ($i=0,\ldots,3$)
were then fitted with quartic polynomial functions of $\sigma_{\ln\rho}$,
as done for the coefficients $c_{i}$.

The final coefficients are shown in Table \ref{table}. Figure \ref{fig2}
shows the resulting $\gamma$'s as functions of $\gamma_{{\rm g}}$
for every $\sigma_{\ln\rho}$. Figure \ref{fig3} presents the resulting
($\gamma$, $\sigma_{\ln N}$) pairs for every ($\gamma_{{\rm g}}$,
$\sigma_{\ln\rho}$) pair. In the figures, the solid and dotted lines
are the best-fit polynomials to reproduce the ($\gamma$, $\sigma_{\ln N}$)
pairs corresponding to constant $\sigma_{\ln\rho}$ and $\gamma_{{\rm g}}$,
respectively.

The power spectra of lognormal density ($\rho)$ fields are usually
shallower than those of the original Gaussian density ($\ln\rho$)
fields, i.e., $\gamma<\gamma_{{\rm g}}$. The property that the power
spectra of $\ln\rho$ are stiffer than those of $\rho$ in turbulence
simulations is also noticeable in Table 2 of \citet{Kowal2007}. The
difference between $\gamma$ and $\gamma_{{\rm g}}$ becomes more
significant for a larger $\sigma_{\ln\rho}$ and for shallower (smaller)
$\gamma_{{\rm g}}$. Column density variance is always smaller than
the variance of 3D density, as expected. The ratio of $\sigma_{\ln N}$
to $\sigma_{\ln\rho}$ becomes smaller for a shallower $\gamma$.
This is caused by less spatial correlation in the density fields with
a shallower spectral slope. For instance, the density fields with
$\gamma=0$ have no spatial correlation.

The equations relating $\gamma$ and $\gamma_{{\rm g}}$ will be useful
in generating the realistic ISM density structures. The equation relating
$\sigma_{\ln N}$ to $\gamma$ and $\sigma_{\ln\rho}$ or Figure \ref{fig3}
provides a tool to derive $\sigma_{\ln\rho}$ from the observable
$\sigma_{\ln N}$ and $\gamma$. Once we obtain $\gamma$ and $\sigma_{\ln N}$
from observations, we can vary $\sigma_{\ln\rho}$ to find the best-fit
value that gives the observed $\sigma_{\ln N}$.

\begin{figure}[t]
\begin{centering}
\includegraphics[scale=0.42]{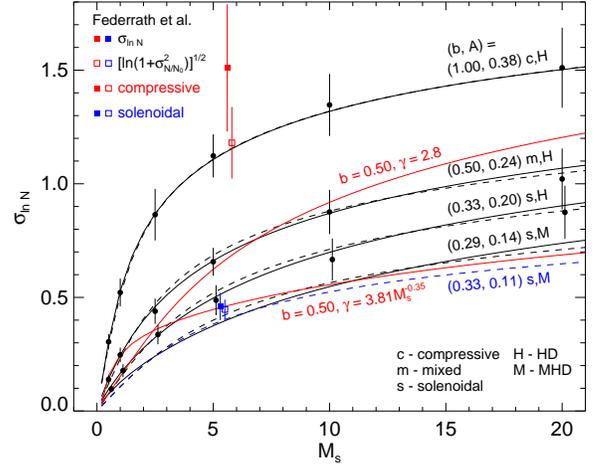}
\par\end{centering}

\caption{\label{fig4}Column density variance-Mach number relationship. The
upper three black-curves represent the cases of compressive ($b=1$),
mixing ($b=0.5$), and solenoidal mode ($b=1/3$). The lowest black
curves are obtained by assuming $b=0.29$ and $\gamma=3.81M_{{\rm s}}^{-0.16}-0.3$
to mimic the magnetic field effects. Dashed black lines represent
the best-fit results to Equation (\ref{eq:1}) with the parameters
($b$, $A$) denoted along the curves. Dashed blue line shows the
curve of \citet{Burkhart2012}. Forcing types (c, m, s) and the simulation
regimes (H, M) are also denoted. Red lines are obtained by varying
the $\gamma-M_{{\rm s}}$ relationship arbitrarily and demonstrate
that Equation (\ref{eq:1}) is not satisfied unless proper relationships
between $\gamma$ and $M_{{\rm s}}$ are assumed.}
\medskip{}
\end{figure}

\section{COLUMN DENSITY VARIANCE-MACH NUMBER RELATIONSHIP}

We now investigate the relationship between $\sigma_{\ln N}$ and
$M_{{\rm s}}$ by using the expressions derived in Section 2. In the
above, we have assumed that $\gamma$ and $\sigma_{\ln\rho}$ are
independent variables. However, $\gamma$ and $\sigma_{\ln\rho}$
are not independent in the real turbulent media, but rather depend
on $M_{{\rm s}}$. If we ignore magnetic fields, the statistical properties
are uniquely determined by $M_{{\rm s}}$ and $b$. We therefore need
to specify the dependencies of $\gamma$ and $\sigma_{\ln\rho}$ on
$M_{{\rm s}}$.

\citet{Seon2009} combined the results of \citet{Kim05}, \citet{Kritsuk06},
and \citet{Padoan04}, and obtained a simple relation between $M_{{\rm s}}$
and $\gamma$, i.e., $\gamma=3.81M_{{\rm s}}^{-0.16}$, which may
be applicable to solenoidal forcing. The power spectra of the compressively
driven turbulence are considerably steeper than those of the solenoidal
forcing. The spectral slopes at $M_{{\rm s}}\approx$ 2.3 and 5.6
for compressive forcing are provided by \citet{Schmidt2009} and \citet{Federrath2009},
respectively, and are $\sim$ 0.6 larger than the values for the solenoidal
case. We therefore assume that $\gamma=3.81M_{{\rm s}}^{-0.16}+0.6$
for the compressive mode. There are two transverse (solenoidal) and
one longitudinal (compressive) wave mode in the 3D space. Therefore,
the solenoidal and compressive forcing would be mixed with a ratio
of 2:1 if no preferential forcing mode were provided \citep{Federrath2010}.
The power spectrum of this natural mixing mode might be obtained by
adding two power spectra for solenoidal and compressive forcing with
a ratio of 2:1. The resulting power spectrum is also represented by
a power-law with a spectral slope, which is at most $\sim0.06$ larger
than the slope of the solenoidal case. We therefore assume the same
spectral slope as the solenoidal forcing case for the natural mixing
case.

The density variance in logarithmic scale is obtained from the variance
in linear scale using the property of a lognormal distribution, i.e.,
$\sigma_{\ln\rho}^{2}=\ln(1+\sigma_{\rho/\rho_{0}}^{2})=\ln(1+b^{2}M_{{\rm s}}^{2})$.
The forcing parameter $b$ is 1/3, 0.5, and 1 for solenoidal, natural
mixing, and compressive modes, respectively \citep{Federrath2008,Federrath2009,Federrath2010}.

Using $\gamma$ and $\sigma_{\ln\rho}$ as functions of $M_{{\rm s}}$,
and Equation (\ref{eq:2}), we calculate $\sigma_{\ln N}$ for solenoidal,
natural mixing, and compressive modes as functions of $M_{{\rm s}}$.
The results are shown in Figure \ref{fig4}, wherein the upper three
black-curves show $\sigma_{\ln N}$ as functions of $M_{{\rm s}}$
for three different forcing types. We fitted the obtained curves with
Equation (\ref{eq:1}). The best-fit $A$ values are found to be 0.2,
0.24, and 0.38 for the solenoidal, natural mixing, and compressive
modes, respectively. In Figure \ref{fig4}, the $b$ and best-fit
$A$ values are denoted in parentheses along the curves. The best-fit
curves are shown as dashed lines.

Applying the expression relating $\gamma$ and $\gamma_{{\rm g}}$
derived with the 128$^{3}$ dataset, we also generated the lognormal
density fields with a higher resolution of 512$^{3}$ for $M_{{\rm s}}=$
0.5, 1, 2.5, 5, 10, and 20. For every combination of parameters ($b$,
$M_{{\rm s}}$), we produced 30 realizations and calculated the ensemble
average of $\sigma_{\ln N}$ and its statistical dispersion. Circles
and error bars in Figure \ref{fig4} represent the ensemble averages
and the dispersions, respectively, which closely follow the curves
calculated with the 128$^{3}$ dataset. We therefore conclude that
the expressions obtained with a rather lower resolution of 128$^{3}$
are also applicable to the higher resolution of 512$^{3}$. The error
bars obtained with the 512$^{3}$ resolution indicate the ranges of
variation between realizations.

In Figure \ref{fig4}, we over-plotted the data points obtained from
the HD simulations with solenoidal (blue squares) and compressive
forcing (red squares) modes of \citet{Federrath2010}, who provide
the standard deviations of linear column density ($\sigma_{N/N_{0}}$)
and logarithmic column density ($\sigma_{\ln N}$). The filled squares
show the $\sigma_{\ln N}$ values that they directly measured with
the logarithmic column density and the open squares represent the
values $[\ln(1+\sigma_{N/N_{0}}^{2})]^{1/2}$ calculated with the
standard deviations of the column density $\sigma_{N/N_{0}}$ assuming
that the column density PDF is a perfect lognormal distribution. The
values estimated with the two different methods show a significant
discrepancy, especially for compressive mode, indicating that the
column density PDF is not a perfect lognormal (see also \citealt{Price2011}
for the discrepancies). Indeed, the PDFs for compressive forcing show
large departures from a perfect lognormal function \citep{Federrath2010}.
Even with this departure from the perfect lognormal function, the
present results are consistent with their results within the variation
ranges denoted by error bars in the figure.

In the above analyses, we have assumed HD cases. \citet{Burkhart2012}
fitted the MHD simulation results for solenoidal forcing with Equation
(\ref{eq:1}) and obtained a best-fit $A$ of 0.11, which is different
from our result of $A=0.2$. The difference might be attributable
to the magnetic field effects. First, the density variance in magnetized
gas is significantly lower than that in the HD counterparts \citep{Ostriker01,Price2011,Molina2012}.
The effect could be mimicked by reducing $b$ below the value of the
HD case, while assuming the same relationship between $\gamma$ and
$M_{{\rm s}}$ for the HD case. We found that the $\sigma_{\ln N}^{2}-M_{{\rm s}}$
relationship derived with $b=0.22$ is very close to the result of
\citet{Burkhart2012}. Second, the spectral slope of the strongly
magnetized media appears to be shallower than those of unmagnetized
or weakly magnetized media. \citet{Padoan04} found that the MHD model
with approximate equipartition of kinetic and magnetic energies yields
a shallower density power spectrum than the super-Alfv{\'e}nic model.
The results of \citet{Kowal2007} also provide some indications that
the power-law slopes of the supersonic, sub-Alfv{\'e}nic models are
slightly smaller than those of super-Alfv{\'e}nic models. Their spectral
slopes for the cases of a strong magnetic field are lower than the
values expected from $\gamma=3.81M_{{\rm s}}^{-0.16}$ by amounts
of $\sim0.3-0.4$. Indeed, we could reproduce the best-fit curve of
\citet{Burkhart2012} with a spectral slope of $\gamma=3.81M_{{\rm s}}^{-0.16}-0.5$
and $b=1/3$. In a real situation, both effects may play roles together.
Therefore, we lowered both $\gamma$ and $b$ and found that the results
with $\gamma=3.81M_{{\rm s}}^{-0.16}-0.3$ and $b=0.29$ match the
result of \citet{Burkhart2012}, as shown by the lowest black curve
in Figure \ref{fig4}. In the figure, the black and blue dashed lines
denote the best-fit results to Equation (\ref{eq:1}) with $b=0.29$
and $b=1/3$, respectively.

\citet{Burkhart2010} presented $\gamma$ as a function of $M_{{\rm s}}$
for MHD simulations with different magnetic field strengths and found
that the magnetic field effect on $\gamma$ is more important at lower
$M_{{\rm s}}$ than at higher $M_{{\rm s}}$. Therefore, we may need
further investigations on the magnetic field effects. 

Another aspect worth noting is that we can constrain the condition
for Equation (\ref{eq:1}) to hold, by varying the equation between
$\gamma$ and $M_{{\rm {\rm s}}}$. For instance, in Figure \ref{fig4},
we plotted the $\sigma_{\ln N}-M_{{\rm s}}$ relationship by assuming
a constant spectral slope of $\gamma=2.8$ (upper red curve) for $b=0.5$.
It is obvious that the resulting curve cannot be expressed by Equation
(\ref{eq:1}). As we assume $\gamma$ increases with $M_{{\rm s}}$,
we obtain more rapidly increasing curves than the red curve. On the
other hand, if we assume $\gamma$ decreases more rapidly than $\gamma=3.81M_{{\rm s}}^{-0.16}$
with increasing $M_{{\rm s}}$, the resulting curve shows a much slower
increase with increasing $M_{{\rm s}}$, as indicated by the lower
red curve obtained for $\gamma\propto M_{{\rm s}}^{-0.35}$. Therefore,
we conclude that the necessary condition for Equation (\ref{eq:1})
is that the spectral slope should moderately decrease as $M_{{\rm s}}$
increases.

\section{CONCLUDING REMARKS}

We investigated the dependency of the column density variance on the
3D density variance and the power-law slope of the density power spectrum
by using the fBm structures. Adopting the spectral slopes and 3D density
variances that are appropriate in the HD regime, we obtained the same
relationship between the column density variance and the Mach number
as suggested by \citet{Burkhart2012}, but with a different scaling
parameter $A$. However, when the magnetic field effects were included,
the parameter $A$ agreed with the MHD result of \citet{Burkhart2012}.
Magnetic fields prevent the turbulent gas from reaching very low densities
as well as very high densities \citep{Molina2012}, and consequently
have important implications for the models of the star formation rate
\citep[e.g., ][]{Federrath2012}.

The approach adopted in this letter is similar to that of \citet{Padoan97}
in the point that they also use random density fields to derive the
3D density variance. The fBm structures also have been used to investigate
the density structure of molecular clouds \citep{Stutzki1998} and
the stellar initial mass function \citep{Elmegreen2002,Shadmehri2012}.
Note that \citet{Stutzki1998} neglected the dispersion in Fourier
amplitudes in generating Gaussian random fields. We compared the cases
of fixed Fourier amplitudes with the results presented here, and found
that the fixed amplitudes give rise to slightly stiffer spectral slopes
than the present results.

The present results can be applied directly to the observations to
infer the 3D density variance from the observed column density variance
and spectral slope. We can vary $\sigma_{\ln\rho}$ in Equations (\ref{eq:2})
and (\ref{eq:3}) to find the best-fit value that matches the observed
$\sigma_{\ln N}$. In applying the results, we need to consider the
effects of instrument noise and telescope smoothing. However, the
inclusion of noise and smoothing may not severely affect the applicability
of the results, as in \citet{Burkhart2012}.

The density fluctuation of the ISM plays a crucial role in understanding
the propagation of radiation. The intensity PDF of the dust-scattered
starlight in far-ultraviolet was found to be lognormal, which might
be caused by the density structure of interstellar dust \citep{Seon2011a}.
Escape of the ionizing radiation field through the clumpy ISM has
been argued to be the origin of ubiquitous diffuse ionized gas \citep[e.g., ][]{Haffner2009}.
This has not, however, been clearly confirmed using realistic ISM
density structures \citep[c.f., ][]{Seon2009,SeonWitt2012}. The present
study provides a practical method to generate the realistic density
structures for the radiative transfer problems.

\end{document}